%% file: paper.tex
\PassOptionsToPackage{dvipsnames}{xcolor}
\documentclass[10pt,screen,sigplan,authorversion]{acmart}

\usepackage[all]{nowidow}
\usepackage{xcolor}
\usepackage{subcaption}
\usepackage{hyperref}
\usepackage{acronym}
\usepackage[htt]{hyphenat}

\usepackage[capitalise,noabbrev]{cleveref}
\crefformat{section}{\S#2#1#3} 
\crefformat{subsection}{\S#2#1#3}
\crefformat{subsubsection}{\S#2#1#3}

\newcommand{\name}{\text{\textsc{ProFaaStinate}}}

\newcommand{\caller}{Call Scheduler}

\begin{document}

\author{Trever Schirmer}
\affiliation{%
    \institution{TU Berlin \& ECDF}
    \city{Berlin}
    \country{Germany}}
\email{ts@mcc.tu-berlin.de}

\author{Natalie Carl}
\affiliation{%
    \institution{TU Berlin \& ECDF}
    \city{Berlin}
    \country{Germany}}
\email{vc@mcc.tu-berlin.de}

\author{Tobias Pfandzelter}
\affiliation{%
    \institution{TU Berlin \& ECDF}
    \city{Berlin}
    \country{Germany}}
\email{tp@mcc.tu-berlin.de}

\author{David Bermbach}
\affiliation{%
    \institution{TU Berlin \& ECDF}
    \city{Berlin}
    \country{Germany}}
\email{db@mcc.tu-berlin.de}

\title[\name{}]{\name{}: Delaying Serverless Function
    Calls to Optimize Platform Performance}



\acmYear{2023}\copyrightyear{2023}
\setcopyright{acmlicensed}
\acmConference[WoSC '23]{9th International Workshop on Serverless Computing}{December 11--15, 2023}{Bologna, Italy}
\acmBooktitle{9th International Workshop on Serverless Computing (WoSC '23), December 11--15, 2023, Bologna, Italy}
\acmPrice{15.00}
\acmDOI{10.1145/3631295.3631393}
\acmISBN{979-8-4007-0455-0/23/12}

\begin{abstract}
    Function-as-a-Service (FaaS) enables developers to run serverless applications without managing operational tasks.
    In current FaaS platforms, both synchronous and asynchronous calls are executed immediately.
    In this paper, we present \name{}, which extends serverless platforms to enable delayed execution of asynchronous function calls.
    This allows platforms to execute calls at convenient times with higher resource availability or lower load.
    \name{} is able to optimize performance without requiring deep integration into the rest of the platform, or a complex systems model.
    In our evaluation, our prototype built on top of Nuclio can reduce request response latency and workflow duration while also preventing the system from being overloaded during load peaks.
    Using a document preparation use case, we show a 54\% reduction in average request response latency.
    This reduction in resource usage benefits both platforms and users as cost savings.
\end{abstract}

\maketitle

\input{sections/1_intro}
\input{sections/2_system}
\input{sections/3_eval}

\input{sections/6_lim_conclusion}
\input{sections/5_related}

\input{sections/7_conclusion}

\bibliographystyle{ACM-Reference-Format}
\balance
\bibliography{bibliography.bib}

\end{document}

%% file: sections/1_intro.tex
\begin{figure*}
    \centering
    \includegraphics[width=\textwidth]{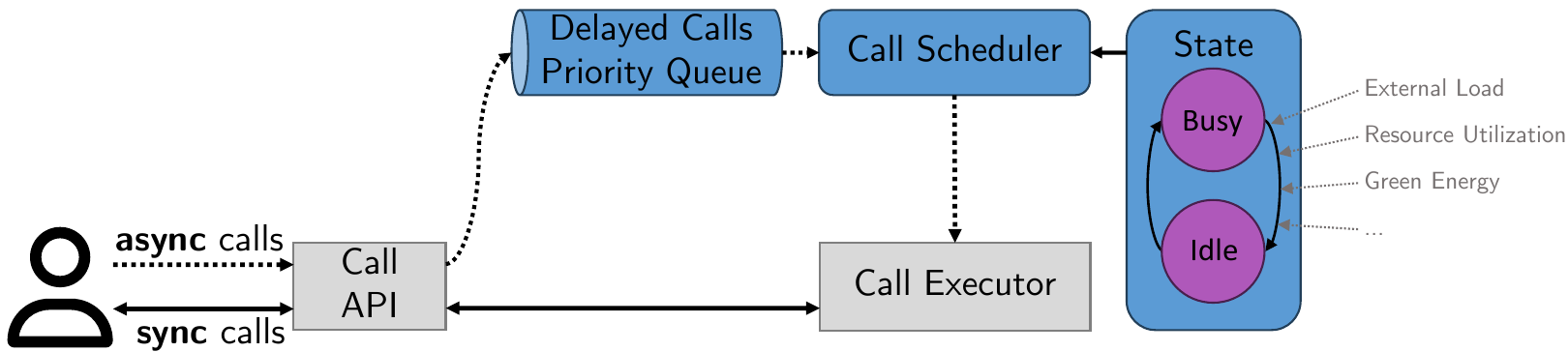}
    \caption{
        A regular FaaS platform contains a frontend call API and a call executor that executes function invocations (both shown in gray).
        \name{} adds a priority queue and call scheduler (shown in blue).
        Incoming asynchronous calls are put into a priority queue, whereas the \caller{} is then responsible for executing delayed calls.
        It has two states, busy and idle, which are influenced by monitoring data.
        In busy mode, only urgent calls are executed.
        In idle mode, urgent and additional non-urgent calls are executed.
    }
    \Description{Users send their asynchronous and synchronous calls to the Call API that already exists. Synchronous calls are routed to the already existing Call Executor. Asynchronous calls are put into a Delayed Calls Priority Queue, from where they are taken by the Call Scheduler, which sends them to the Call Executor. The Call Scheduler has two States, Idle and Busy, which are influenced  by External Load, Resource Utilization, Green Energy, \dots}
    \label{fig:arch}
\end{figure*}

\section{Introduction}
\label{sec:introduction}
With Function-as-a-Service (FaaS), all operational tasks are managed by the serverless platform, enabling developers to focus on writing code and increasing their productivity~\cite{,paper_bermbach2021_cloud_engineering}.
FaaS is a popular cloud execution model, with offerings by all major cloud providers, e.g., Google Cloud Functions\footnote{\url{https://cloud.google.com/functions/}} and Amazon Web Services Lambda\footnote{\url{https://aws.amazon.com/lambda/}}, where function calls are billed on a pay-per-use basis at millisecond granularity~\cite{Hendrickson2016-pw,Castro_2019}.
This enables elastic and scalable applications comprising multiple event-driven stateless functions~\cite{Hendrickson2016-pw,Eismann_2021_Why, paper_bermbach2021_cloud_engineering}.
FaaS is also popular for local computing use cases, with companies running private FaaS platforms inside their data centers, mostly running on top of Kubernetes~\cite{Eismann_2021_Review,Rensin_2015_KubernetesWhitepaper}.
In High Performance Computing, FaaS can be used to abstract from complex operational management of supercomputers~\cite{Spillner_2018_ServerlessHPC,Chard_2020_funcx}.

FaaS functions can be invoked in two different ways: synchronously (i.e., the calling component waits for the result to continue operation) or asynchronously (i.e., the calling component does not wait for the result).
Synchronous calls are often used for implementing web APIs,
whereas asynchronous invocations may be used for processing event streams~\cite{Eismann_2021_3c}.
In current FaaS platforms, both types of function calls are to our knowledge treated the same and executed immediately.

This immediate execution is necessary for synchronous calls as the calling component waits for the result.
However, when calling a function asynchronously, it might be acceptable to delay the execution of the function to a later point.
Eismann et al.~show that $\geq{}\!30\%$ of serverless applications have no latency requirement~\cite{Eismann_2021_3c}.
Giving the serverless platform the ability to delay the execution of calls (up to a certain latency objective) could thus increase scheduling flexibility, reducing cost and overall resource consumption.

The main idea of \name{} is to delay incoming asynchronous calls when the platform is resource-limited and to execute delayed calls when the platform has excess resources.
What constitutes a resource limitation depends on underlying infrastructure, business model, and scale:
We have shown in previous work that Google Cloud Functions (GCF) has a diurnal performance variability of up to 14\%~\cite{schirmer2023nightshift}.
This performance variability also can lead to a 14\% cost reduction due to the pay-per-use billing model so that delaying the execution until the night can decrease cost.
Similarly, the concept of `spot' instances in cloud computing shows that cloud providers have varying request load on their infrastructure~\cite{chohan_2010_spot}.
Smaller-scale FaaS deployments running in local clouds might be resource constrained during times of high load, as they are contending with other workloads for limited resources.
Delaying execution of functions also allows platforms to batch calls to functions with low utilization, for which every call would lead to a cold start if it were executed immediately.
Cold starts happen when a new function instance needs to be started to handle a request~\cite{paper_bermbach2020_faas_coldstarts,Manner_2018_Coldstarts}.
They have higher latency and cost than executing calls on warm, already running instances.

Our intuition is that simply delaying asynchronous function execution during times of resource limitation to periods of low load can spread resource usage over a longer time and thereby decrease the observable latency of synchronous requests.

Crucially, this requires neither an advanced systems model, complex scheduling mechanisms, nor predicting platform load.
In this paper, we present \name{}, a system to exploit load and resource availability fluctuations in FaaS platforms by allowing the platform to delay the execution of functions to a later time when they can be executed faster.

We make the following main contributions:

\begin{itemize}
    \item We describe considerations for delayed execution of serverless functions and present an architecture to extend existing serverless platforms (\cref{sec:system}).
    \item We evaluate our proposal by implementing it on top of the Nuclio\footnote{\url{https://nuclio.io}} open-source serverless platform (\cref{sec:eval}).
    \item We discuss the limitations of our system and future research avenues (\cref{sec:lim}).
\end{itemize}

%% file: sections/2_system.tex
\section{\name{}: Delaying Execution of Function Calls}
\label{sec:system}
As we show in~\cref{fig:arch}, \name{} extends existing FaaS platforms by enabling them to `re-route' asynchronous invocations.
This minimizes the integration into the platform so that the architecture can be used to extend different platforms.
If an incoming call is synchronous, it takes the normal path through the FaaS platform: After arriving at the public call API (i.e., the component that receives calls from users), they are immediately executed by the call executor (i.e., the component that distributes incoming calls to function instances).
\name{} extends the call API with one alternative branch:
Asynchronous invocations are enqueued into a priority queue with a developer-specified latency objective.
This queue is then read by a Call Scheduler, which executes delayed calls using the call executor of the platform.

The Call Scheduler can be in two states that change the amount of calls it sends to the call executor.
Either it has free capacity (idle state), or it does not have excess capacity (busy state).
In busy state, the platform is using up most or all available resources with synchronous calls.
To limit additional resource consumption, the Call Scheduler should only execute delayed calls whose deadline is approaching.
In idle state, the platform has more resources available than are currently consumed by incoming synchronous calls.
This means the Call Scheduler should execute more than only those calls whose deadline is approaching, instead also executing calls with a deadline further in the future.

Which state the Call Scheduler should be in depends on current monitoring data and deployment goals.
The system can be used to minimize cost by delaying calls when resources are slow or expensive.
It is also possible to optimize for other goals, such as minimizing the carbon impact of workloads by delaying execution until sufficient renewable energy is available.

\name{} still has a simple development model: developers can choose to specify the maximum additional delay of functions during their deployment.
The platform is then responsible for delaying invocations and executing them later.

\begin{figure*}
    \centering
    \includegraphics[width=\textwidth]{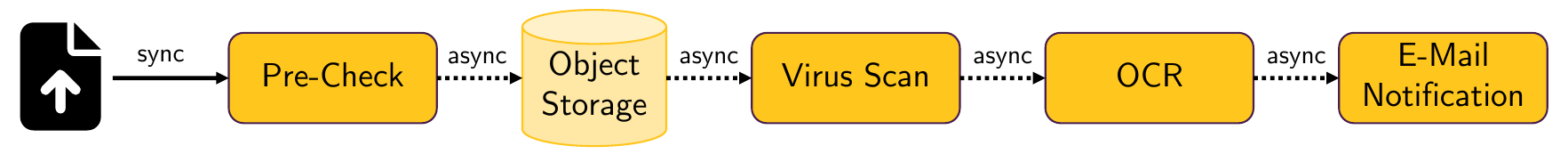}
    \caption{
        Overview of the document preparation use case used in the evaluation. Users synchronously call the Pre-Check function, which checks the PDF for correctness and puts it into object storage.
        From there, a virus scan, optical character recognition (OCR), and an e-mail notification function are called asynchronously.
    }
    \Description{The document upload triggers the Pre-Check function synchronously, which then puts the file into object storage asynchronously. After the Object Storage is an asynchronous call chain from Virus Scan to OCR to E-Mail Notification}
    \label{fig:usecase}
\end{figure*}

%% file: sections/3_eval.tex
\section{Evaluation}
\label{sec:eval}

To evaluate \name{}, we demonstrate that it can save cost and reduce resource consumption in a realistic use case.
Our focus is to demonstrate that it can be used to `shave-off' load peaks by delaying execution when the system is overloaded.
In the following sections, we describe our prototype implementation on top of the Nuclio open-source FaaS platform (\cref{sec:impl}), a possible use case (\cref{sec:useCase}), and experiment design (\cref{sec:expDesign}).
We then present the results of our experiments in \Cref{sec:results}.
We make our artifacts available as open-source software.\footnote{\url{https://github.com/umbrellerde/nuclio/tree/1.11.x}}

\subsection{Implementation}
\label{sec:impl}

We implement a prototype for \name{} by extending the open-source serverless platform Nuclio.\footnote{\url{https://nuclio.io}}
First, we check every incoming function invocation for whether it is asynchronous.
Asynchronous calls are accepted for execution (i.e., a \texttt{204} HTTP response is sent), serialized, and persisted to a database.
The second component (\caller{}) is responsible for executing delayed calls using the scheduling rules described in \Cref{sec:system}.
To execute calls, the \caller{} uses the normal synchronous invocation API offered by Nuclio, so that the request is executed immediately.

Our prototype changes state between idle and busy depending on the amount of free CPU resources available to functions.
This information is available out of band from the serverless platform by collecting metrics from the underlying container orchestrator (Docker or Kubernetes).
For the purposes of this evaluation, the \caller{} changes its state to busy if the average CPU utilization of function runtimes is ${\geq{}}90\!\%$ for 30 second, and to idle if the utilization is $\leq{}\!60\%$ for 30 seconds.

\subsection{Use Case}
\label{sec:useCase}

In our experiments, we aim to show that \name{} can be used in resource-intensive applications which, however, are not time-sensitive, so execution can be delayed.
To realize this, we focus on a document preparation use case.
We implement this as a stream application, a popular serverless use case~\cite{Eismann_2021_3c}.
\Cref{fig:usecase} shows an overview of the functions in this use case.
In a first step, users upload a scanned document.
After a quick pre-check of the document to give users immediate feedback, the document is put into object storage.
This asynchronously triggers a virus scan, the completion of which then triggers optical character recognition (OCR).
Afterwards, users are informed via e-mail that their document has been processed.
The pre-check needs to happen synchronously to immediately inform users of any obvious errors.
All other functions can be delayed by \name{}, as the results are not required immediately.

\subsection{Experiment Design}
\label{sec:expDesign}

We demonstrate the feasibility of \name{} and investigate the behavior of our proof-of-concept prototype using an implementation of the described use case with simulated users.
To simulate many users concurrently scanning and uploading documents, we put the system under a constant load of one document uploaded per second for 30 minutes.
In parallel, we put the CPU in the system under test under an artificial load to simulate a load peak of other workloads running on the system.
By using \name{}, the CPU-intensive tasks can be delayed to a later time so that the node is not overwhelmed during peak load and can perform the pre-check that needs to happen synchronously faster.
The experiment is split into three phases:
During the \emph{load peak} phase in the first ten minutes of the experiment, the artificial CPU load it is set to 80\% (to simulate other workloads using up almost all resources).
During the \emph{cooldown} phase, the CPU load linearly decreases over ten minutes to 15\%.
In the following \emph{low load} phase, the CPU load stays at 15\% for another ten minutes (to simulate most other workloads being finished).
This behavior is not atypical for a FaaS platform, whether public in the cloud or in a private data center, as infrastructure load will change, e.g., during the workday~\cite{schirmer2023nightshift}.

We compare two execution models:
As a baseline, all invocations (synchronous and asynchronous) are executed immediately.
We then use \name{} to delay asynchronous invocations where desirable.
We allow a delay of up to seven minutes for the virus scan and OCR functions and a three-minute objective for the e-mail function.
This allows \name{} to delay the majority of complex calls until the load peak is finished.

We deploy our experiments on a Google Cloud Platform \texttt{e2-highmem-8} virtual machine with eight vCPUs and 64GB of memory.

\subsection{Results}
\label{sec:results}

We report three metrics: the CPU utilization, the request-response latency of the synchronous Pre-Check function call for users, and the duration of the whole workflow.

\begin{figure}
    \centering
    \includegraphics[width=0.5\textwidth]{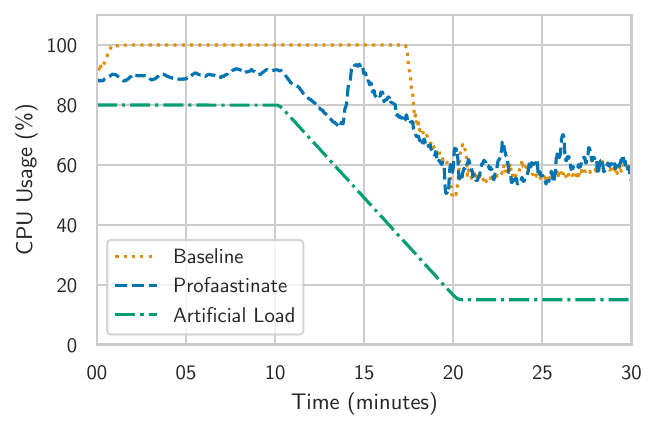}
    \caption{
        CPU usage graph of both experiments. During the load peak (0–10 minutes), \name{} reduces the system strain from 100\% (baseline) to a mean 89\%.
        After the load peak is over, executing delayed invocations leads to a slightly higher CPU load with \name{}.
    }
    \Description{Line plot with Time in minutes on x-axis and CPU Usage in percent on Y-Axis. Baseline has a 100\% PU Usage during the first 15 minutes, which then decreases to around 60\% for the rest of the experiment. Profaastinate has a maximum usage of around 90\%, which then also decreases to a slightly higher line than the baseline experiment.}
    \label{fig:res:cpu}
\end{figure}

\subsubsection*{CPU Utilization}
\Cref{fig:res:cpu} shows the CPU utilization for the duration of our experiment.
During the load peak phase in the baseline experiment, the system is overloaded and using all CPUs (80\% mean utilization).
This increases the request-response latency and workflow duration of all calls made during that time.
With \name{}, the CPU is not overloaded during the load peak phase (89\% utilization, 9\%pt. over artificial load).
After the load peak is over, \name{} switches over to idle mode, where it executes additional delayed requests.
This slightly increases the average CPU load during the low load phase to 59\% compared to 57\% (baseline).

Note that there is a load spike for \name{} at 14 minutes:
This is not caused by CPU utilization falling below 80\% and the Call Scheduler executing queued requests (recall that this only occurs once 60\% utilization is reached).
Rather, this is the result of asynchronous invocations for the OCR function reaching their deadline.
During busy state, a workflow started during the beginning of the experiment will have a deadline for its virus scan function at the seven-minute mark, which will then set the OCR deadline to 14 minutes (from workflow start).

\begin{figure}
    \centering
    \includegraphics[width=0.5\textwidth]{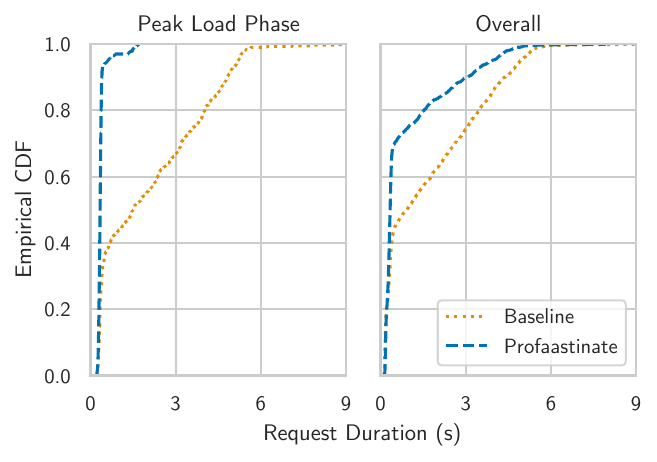}
    \caption{
        Request-response latency during the load peak phase and the overall in both experiments.
        In both cases, \name{} reduces request-response latency compared to baseline, yet the effect is more pronounced during peak load.
        Overall, the fastest 50\% of calls have a similar request response latency in both experiments, while the slower 50\% are faster when using \name{}.
    }
    \Description{Two ECDF plots showing the Request Duration distribution. On the left side, the load peak phase is shown. On the right side, the whole experiment is shown. The fastest 40\% of calls have a similarly low ($\leq 1$s) duration. Afterwards, the \name{} line is further to the left than the baseline line.}
    \label{fig:res:reqEcdf}
\end{figure}

\subsubsection*{Request-Response Latency}

\Cref{fig:res:reqEcdf} shows the distribution of request-response latency from a client perspective, i.e., the latency perceived by users.

During the load peak, resource contention between synchronous, client-facing functions and resource-hungry asynchronous processing functions leads to a higher request duration for 50\% of requests in the baseline.
\name{} consistently leads to a fast execution (standard deviation 1.8s for the baseline and 0.2s for \name{}) and reduces peak latency (99\textsuperscript{th} percentile latency reduced from 5.6s (baseline) to 1.5s (\name{})).

As this effect is more pronounced during the peak load phase, the results clearly show how simply delaying resource-intensive function execution with \name{} helps improve perceived system performance.

\begin{figure}
    \centering
    \includegraphics[width=0.5\textwidth]{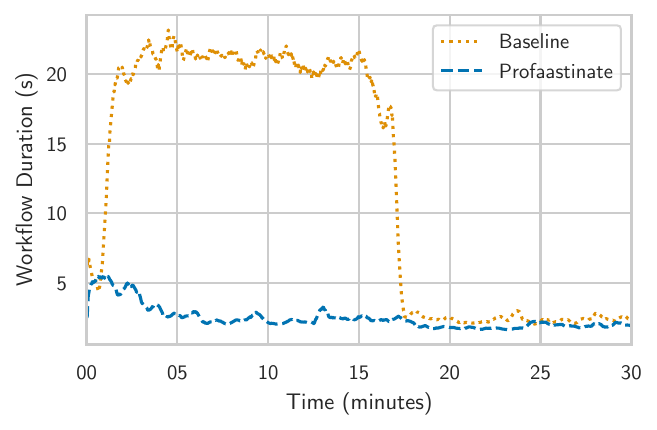}
    \caption{
        Workflow Duration during the experiments.
        While the CPU is at 100\% utilization, the workflow duration is very high in the baseline.
        In \name{}, the workflow duration is lower during the load peak.
    }
    \label{fig:res:worfklow}
\end{figure}

\subsubsection*{Workflow Duration}

We define `workflow duration' as the sum of execution durations of all functions involved in a single document processing request.
The results in \cref{fig:res:worfklow} again show the impact of resource contention during the load peak phase of the experiment.
In our baseline, all functions are executed as soon as they are called, leading to an average workflow duration of 19s during the load peak phase and 2.3s during low load phase.
With \name{}, the actual execution of asynchronous calls can happen after the load peak, leading to fewer stragglers and a similar average workflow duration (99\textsuperscript{th} percentile: 6.3s, mean 2.4s).

Note that for requests started during the start of the load peak phase, workflow duration is slightly increased as the first resource-intensive OCR functions are executed during the load peak phase.

%% file: sections/6_lim_conclusion.tex
\section{Limitations \& Future Work}
\label{sec:lim}

By simply delaying asynchronous function execution during times of high load, \name{} improves FaaS system utilization and helps allocate resources to more important synchronous functions.
We plan on building upon this work by further integrating serverless platforms and applications.

\subsubsection*{Scheduler}

Currently, \name{} has a simple scheduling mechanism, which can already improve performance.
There is a trade-off when deciding on the scheduler complexity as a more complex scheduler might further improve performance but also might use more resources and be more prone to over-fitting.
As one example, our scheduler only looks at the deadline of calls to decide which calls to execute next.
A more complex scheduler might also group calls to one function together to limit cold starts.
Our system is extensible to use different schedulers, so that we can research this in future work.

\subsubsection*{Workflows}

In \name{}, developers have to decide on the maximum additional delay per function.
This is relatively trivial for workflows comprising only one function, but for more complex workflows it would be easier to decide when the last function needs to be finished instead.
In future work, we plan to enable this by integrating \name{} into our framework \textsc{Fusionize}~\cite{paper_schirmer2022_fusionize}, which already generates the workflow graph from monitoring data.

%% file: sections/5_related.tex
\section{Related Work}
\label{sec:related}

The increasing research interest in serverless computing has led to several approaches aiming to improve the performance of singular FaaS functions:
This can be achieved by reducing the number of cold starts~\cite{Bardsley_2018_coldstarts,paper_bermbach2020_faas_coldstarts}, by optimizing the infrastructure configuration~\cite{Cordingly_2022_memorysizes}, or using hardware acceleration~\cite{Werner_2022_Hardless,Pemberton_2022_KaaS}.
Other research proposes to improve performance and reduce cost by executing some functions calls outside the serverless platform, e.g., using Container-as-a-Service~\cite{Czentye_2019} or VMs~\cite{Horovitz_2019_FaaStest,Jain_2020_SplitServe} as backends.
These approaches can be used in conjunction with \name{}, as they focus on different aspects of the function invocation lifecycle.

Other work focuses on how to automatically split up serverless applications into different functions~\cite{paper_schirmer2022_fusionize,Mahgoub_2022_WISEFUSE}, which could be integrated with \name{} to also automate the configuration of the allowed latency.

Scheduling of function invocations has also received considerable attention:
With \emph{Ensure}, Suresh et al.~\cite{Suresh_2020_ENSURE} present a scheduler that optimizes placement of function calls on infrastructure and automates scaling.
The goal of this approach is the minimization of delay between invocation and execution.
We argue that this is relevant primarily for synchronous execution — it is feasible to view this as a complementary solution to \name{}.
Zhang et al.~\cite{Zhang_2021_Caerus} present a scheduler for serverless multi-stage workflows, with a focus on data analytics.
In their work, they optimize the task-level schedule given an existing execution plan.
Compared to their work, \name{} does not need an execution plan as it enqueues requests as they come in.

%% file: sections/7_conclusion.tex
\section{Conclusion}
\label{sec:concl}

In this paper, we have presented \name{}, a system for optimizing serverless platform performance by delaying execution of asynchronous function calls.
We have presented the architecture of our system, which only requires shallow integration into the serverless platform.
Our evaluation using a proof-of-concept prototype and file processing use case shows that using a relatively simple scheduling rule can already improve performance during load peaks by delaying calls when available resources are scarce.
In future work, we plan on further integrating serverless applications and platforms.